\documentclass[prb,aps,twocolumn,superscriptaddress,longbibliography]{revtex4-2}
\usepackage{graphicx}
\usepackage{dcolumn}
\usepackage{bm}
\usepackage{dsfont}
\usepackage{lipsum}
\usepackage{ragged2e}
\usepackage{braket}
\usepackage{bbm}
\usepackage{color}
\usepackage{xcolor}
\usepackage{amsfonts}
\usepackage{amsmath}
\usepackage{amsmath,amssymb,latexsym}
\usepackage{wasysym,stmaryrd,marvosym}
\usepackage{mathrsfs}
\usepackage{natbib}
\usepackage{soul}
\usepackage{float} % para H en figuras
\providecommand{\keywords}[1]{\textbf{\textit{Index terms---}} #1}
\usepackage[mathlines]{lineno}
\usepackage[colorinlistoftodos]{todonotes}
\usepackage[colorlinks=true, allcolors=blue]{hyperref}
\begin{document}

\title{Probing quantum geometric capacitance via photonic spin Hall effect}

\author{Yahir Fern\'andez-M\'endez}
\affiliation{Centro de Investigaci\'on Cient\'ifica y de Educaci\'on Superior de Ensenada, Baja California, Apartado Postal 360, 22860, Ensenada, Baja California, M\'exico.}

\author{Ram\'on Carrillo-Bastos}
%\affiliation{Centro de Investigaci\'on Cient\'ifica y de Educaci\'on Superior de Ensenada, Baja California, Apartado Postal 360, 22860, Ensenada, Baja California, M\'exico.}
\affiliation{Facultad de Ciencias, Universidad Autónoma de Baja California, Apartado Postal 1880, 22800 Ensenada, Baja California, México}%

\author{Jes\'us A. Maytorena}
\email{jesusm@ens.cnyn.unam.mx}
\affiliation{Departamento de Física, Centro de Nanociencias y Nanotecnolog\'ia, Universidad
Nacional Aut\'onoma de M\'exico, Apartado Postal 14, 22800 Ensenada, B.C., M\'exico}

\date{\today}

\begin{abstract}

The non-dissipative quasistatic longitudinal optical response of insulators is characterized by an intrinsic geometric capacitance, determined by the ratio of the quantum metric to the energy gap, as recently stablished. We study the photonic spin Hall effect (PSHE) in this low-frequency regime, induced by reflection from an atomically thin material serving
as the interface between two dielectrics. As a function of the magnitude of this geometric capacitance, covering values typical of graphene-family systems and for incidence close to the Brewster angle, the calculated in-plane and transverse beam shifts become comparable to the wavelength and thus detectable. At sufficiently low frequencies,  we derive linear and quadratic approximations that allow the geometric capacitance to be extracted from measurements of the beam displacements. We provide an upper bound for the beam shifts. The results suggest that THz PSHE might be a useful probe of the quantum geometric capacitance of two-dimensional insulators. 

\end{abstract}
\maketitle

Bloch states are characterized not only by the energy dispersion of the bands, but also by the geometric and topological structure of the space of quantum states\cite{NatSciRev2025}. These geometric properties constitute a fundamental framework for understanding topological phases and the physical responses of quantum systems, a connection that remains under intense exploration.\cite{TormaPRL,BernevigQueiroz,Measurement_Kang2024,Queiroz_revisiting,QueirozStepResponse,QueirozInstantaneousResponse}. 
Recently, a new connection was identified \cite{Komissarov}. 
It was shown that
in band insulators the longitudinal optical response at photon energies below the gap is geometric in origin,
determined by the quantum metric weighted by inverse energy differences. This non-dissipative quasistatic response
is originated from coherent, virtual interband transitions between the full valence band and empty conduction band.
It was obtained from Kubo formula that the longitudinal optical conductivities become purely imaginary, 
$\sigma_{ii}=-i\omega\chi_{ii}^0$,
where the static electric susceptibility $\chi^0_{ii}$, also refered to as the intrinsic capacitance of the insulator, is given by
\begin{equation} \label{static2}
\chi_{ii}^0=e^2\sum_{\underset{m\neq n}{n,m}}\sum_{\bf k}(f_{n{\bf k}}-f_{m{\bf k}})\frac{[g^{ii}({\bf k})]_{nm}}{\varepsilon_{m{\bf k}}-\varepsilon_{n{\bf k}}}\,,
\end{equation}
with $n$ being the band index, $f_{n{\bf k}}$ the occupation factor of the $n$-th band with eigenenergy $\varepsilon_{n{\bf k}}$, and $[g^{ii}({\bf k})]_{nm}$ interband matrix elements of the diagonal components of the quantum metric tensor; 
the sums are over all bands and on the momentum ${\bf k}$  defined in the Brillouin zone. 
In reference \cite{Komissarov}, the authors present several examples to
discuss how a measurement of the intrinsic capacitance allows an estimate of the ground state quantum metric. 
On the other hand, the understanding and measurement of the optical response of two-dimensional (2D) materials in the terahertz (THz) and mid-infrared range of frequencies are relevant for potential applications in fields like nano-photonics \cite{polaritonics_vdW} and nano-plasmonics \cite{Low_Avouris}. Usually the focus is
on the intraband response \cite{Lloyd-Hughes_2012,Lloyd-Hughes_2014,Low_Avouris}, however in the case of band insulators for low enough frequencies the role of (virtual) interband transitions should instead be considered \cite{Komissarov,modeling}.

Here we explore the photonic spin Hall effect (PSHE)
at low frequency (THz regime)
in a generic two-dimensional insulator o semiconductor with an optical conductivity determined by the
geometric static susceptibility in Eq.\,(\ref{static2}). In the relevant range of values of $\chi^0_{ii}$ of a graphene-like family, we found that PSHE lies in the measurable range, 
becoming therefore a sensitive probe of the quantum geometric capacitance. We also discuss how
this capacitance could be estimated from PSHE experimental values, and an
upper bound for the spin separation of reflected light.

\begin{figure}[H]
\centering
\includegraphics[width=.45 \textwidth]{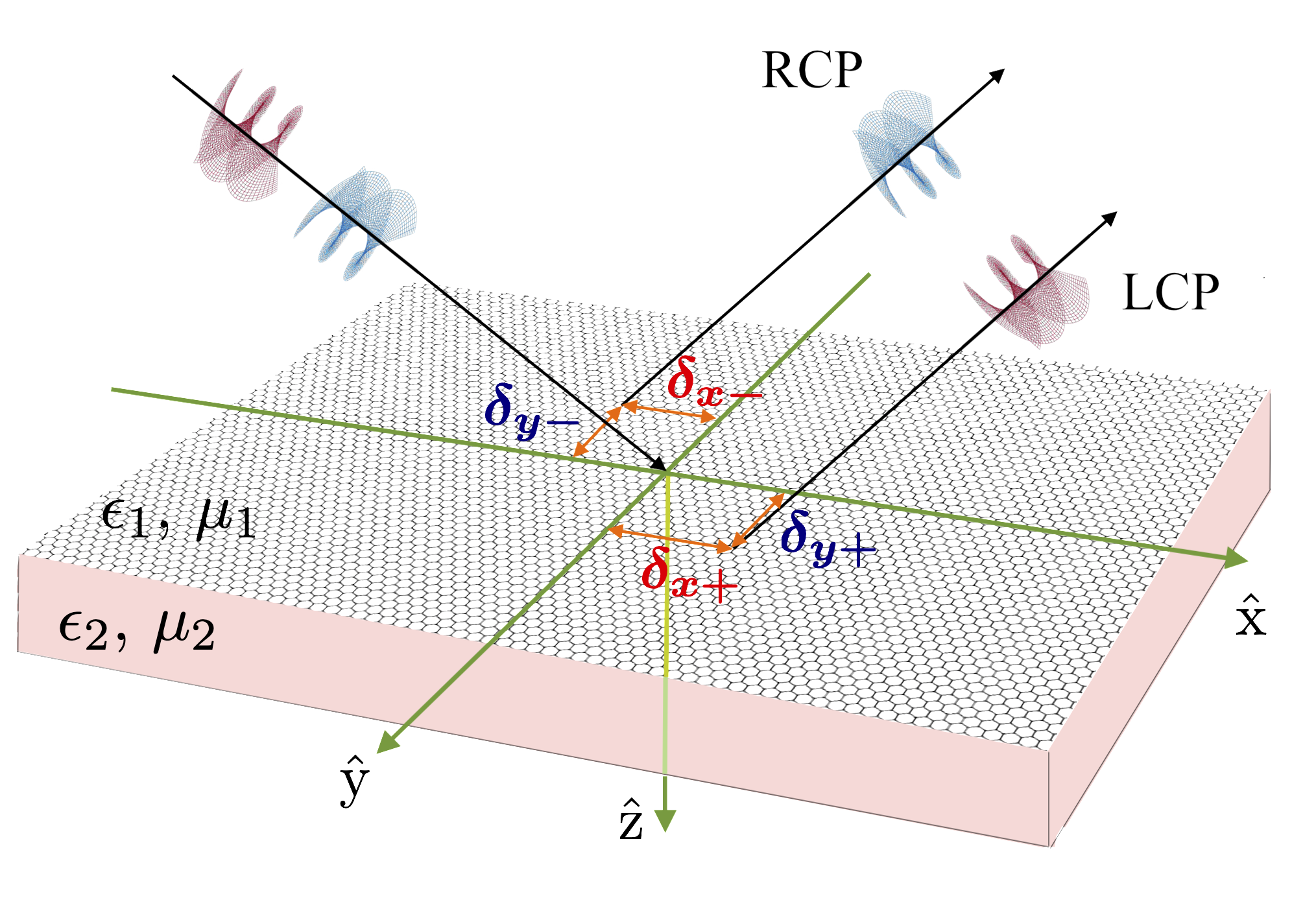}
\caption{Illustration of the PSHE in a 2D system at the interface between
two dielectrics (with optical
constants $\epsilon_1, \mu_1$ and $\epsilon_2, \mu_2$). The in-plane and transverse
intensity centroids shifts, $\delta_{x\pm}$ and $\delta_{y\pm}$, of the 
the left (LCP) and right (RCP) circularly polarized reflected beams are indicated.
The optical response of the 2D medium is characterized by the longitudinal
conductivities $\sigma_{xx}(\omega)$ and $\sigma_{yy}(\omega)$.}
\label{fig:Fig1}
\end{figure}

As a photonic analogue of the electronic spin Hall effect \cite{Hosten,Bliokh2006,Bliokh2007}, the PSHE refers to the
opposite lateral shifts of the intensity distribution centroids of the two spin components of
a light beam when it is reflected or refracted on the interface between two transparent media
(see Fig.\,\ref{fig:Fig1}). 
The physical basis behind this phenomenon is the spin-orbit interaction of light, where the constraint of transverse polarization leads each plane wave component to acquire a distinct geometric phase, thereby coupling momentum and helicity and causing the beam to split into components with oppositely shifted intensity centroids.
Because of its potential applications, this phenomenon has been studied in a wide variety of physical systems, 
such as graphene \cite{Zhou_2012-2,Cai_2017}, 2D semi-Dirac materials \cite{Huang21},
topological insulators \cite{Zhou_2013}, black phosphorus in the mid-IR \cite{anisotropic} and THz \cite{PSHE_BPh} regions, and recently in 2D black arsenic \cite{PSHE_BlackAs},  Haldane materials \cite{PSHE_Haldane}, to mention a few involving 2D materials.
Notable examples of applications include metrological precision \cite{Chen_2020}, 
optical sensing \cite{NanophotonicResonator}, and probing of topological phase transitions 
via THz PSHE in atomically thin 2D semiconductors, including silicene, germanene, and stanene, 
\cite{MShah,Kort-Kamp}.

The theory of beam shifts \cite{Review_JPhysD,Impact,Qin2011} 
by reflection at a flat interface made of a 2D system, separating two
homogeneous media with optical constants $\epsilon_1,\mu_1$ and $\epsilon_2$, $\mu_2$, yields 
\begin{eqnarray}
\delta^{H(V)}_{x\pm}&=&\frac{1}{k_0}\frac{\text{Im}[\partial\ln r_{p(s)}/\partial\theta_i]}{\Lambda_{p(s)}}, 
\label{xHVpm} \\
\delta^{H(V)}_{y\pm}&=&\pm \frac{\cot\theta_i}{k_{0}}\frac{1+\text{Re}[r_{s(p)}/r_{p(s)}]}{\Lambda_{p(s)}},
\label{yHVpm}
\end{eqnarray}
where
\begin{equation} \label{denominator}
 \Lambda_{p(s)}=  1 + \frac{1}{k_{0}^{2} w_{0}^{2}} \left( \left| \frac{\partial \ln r_{p(s)}}{\partial \theta_{i}} \right|^{2} + \left| 1 + \frac{r_{s(p)}}{r_{p(s)}} \right|^{2} \cot^{2} \theta_{i} \right) .
\end{equation}

The spatial displacement $\delta^{H(V)}_{i\pm}$ denotes the in-plane ($i=x$) or transverse ($i=y$) intensity centroid of the right (-) or left (+) circular polarized reflected beam, for an incident beam with horizontal ($H$) or  vertical ($V$) polarization.
The theory assumes an incident monochromatic Gaussian beam of frequency $\omega$, characterized by the radius of its waist $w_0$ and the beam divergence half-angle $\theta_0=2/k_0w_0$ \cite{Kogelnik}, where $k_0=\omega/c=2\pi/\lambda$ is the free space wave number of a wave with wavelength $\lambda$, and a
paraxial approximation corrected up to first-order derivatives.
Expressions in Eqs. (\ref{xHVpm}) and (\ref{yHVpm}) involve the complex Fresnel amplitude
$r_{p(s)}$ for $p$ ($s$) polarization, 
\begin{eqnarray}
r_p&=& \frac{\epsilon_{2} k_{iz} - \epsilon_{1} k_{tz} + 4\pi (k_{iz} k_{tz}/k_0) [\sigma_{xx}(\omega)/c]}{\epsilon_{2} k_{iz} + \epsilon_{1} k_{tz} + 4\pi (k_{iz} k_{tz}/k_0) [\sigma_{xx}(\omega)/c]},
    \label{rpp}\\
r_s&=&\frac{\mu_{2} k_{iz} - \mu_{1} k_{tz} - 4\pi \mu_{1} \mu_{2} k_{0}[\sigma_{yy}(\omega)/c]}{\mu_{2} k_{iz} + \mu_{1} k_{tz} + 4\pi \mu_{1} \mu_{2} k_{0} [\sigma_{yy}(\omega)/c]}, \label{rss}
\end{eqnarray}
and its derivative, evaluated at the central wave vector component of incidence angle $\theta_i$.
The wave numbers $k_{iz} = k_{0} \sqrt{\epsilon_{1}\mu_{1}} \cos \theta_{i}$ and
$k_{tz} = k_{0} \sqrt{\epsilon_{2}\mu_{2}  - \epsilon_{1}\mu_{1}  \sin^{2} \theta_{i}}$ are the
 normal-to-the-surface components of the incident and transmitted central wave vector, with respect to
 the laboratory coordinate system $x,y,z$, with $z=0$ and $y=0$ defining the plane of the interface
 and the incidence plane, respectively. We consider that the response of the atomically thin crystal is 
 described by the longitudinal conductivities $\sigma_{xx}(\omega)$ and $\sigma_{yy}(\omega)$ only.
 Therefore, $p\to s$ and $s\to p$ polarization conversion contributions are absent in formulas
 in Eqs.\,(\ref{xHVpm}) and (\ref{yHVpm}). This further implies that $\delta^{H(V)}_{x+}=
 \delta^{H(V)}_{x-}\equiv \delta^{H(V)}_x$, and $\delta^{H(V)}_{y+}=-\delta^{H(V)}_{y-}\equiv \delta^{H(V)}_y$. 
 The second and third terms in Eq.\,(\ref{denominator}) are $1/k_0^2$ corrections which become relevant for incidence close to the Brewster angle \cite{Qin2011,KongAPL2012}, in particular, the second term is a contribution due to the in-plane spread of wave vectors \cite{Impact}.
 Note from Eq.\,(\ref{xHVpm}) that $\delta^{H(V)}_{x\pm}\propto\partial\text{arg}(r_{p(s)})/\partial\theta_i$,
and therefore that it vanishes when $\sigma_{ii}=0$. 

In the following, we use the low-frequency conductivity $\sigma_{ii}=-i\omega\chi^0_{ii}$ in Eqs.(\ref{rpp}) and (\ref{rss}) to evaluate the in-plane and transverse displacements $\delta^H_x$ and $\delta^H_y$ 
 as functions of the incidence angle in the relevant range of the susceptibility 
 of several 2D systems. For the sake of simplicity, we will focus in results for horizontal polarization only,
 assume $\sigma_{xx}=\sigma_{yy}$, $\epsilon_1=1$, $\epsilon_2=2.3$, $\mu_1=\mu_2=1$, $w_0=30\lambda$
 ($\theta_0\approx 10^{-2}$)
 and abbreviate $\chi^0_{xx}=\chi^0_{yy}\equiv \chi_0$. 
 Note that we are using Gaussian units, therefore the conductivity of the 2D system has units of velocity
 and the susceptibility of length, such that $\sigma_{ii}/c=-ik_0\chi^0_{ii}$ is dimensionless.

\begin{figure}[H]
\centering
\includegraphics[width=.47 \textwidth]{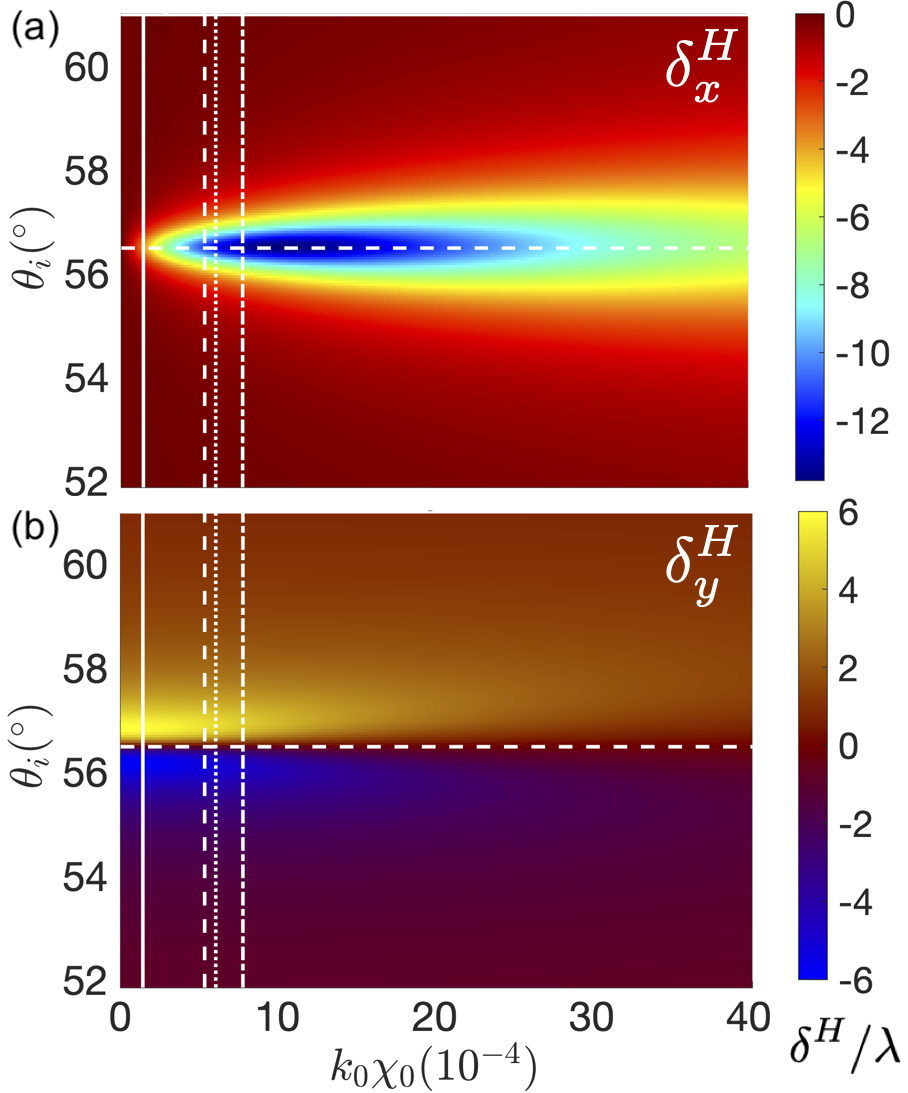}
\caption{In-plane and transverse beam shifts (a) $\delta^H_x$ and (b) $\delta^H_y$ of the LCP 
reflected beam as functions of the
normalized static susceptibility $k_0\chi_0$ and the incidence angle $\theta_i$,
around the Brewster angle $\theta_B=\arctan(\sqrt{\epsilon_2/\epsilon_1})$ (horizontal
dashed line). An incident Gaussian beam with horizontal polarization is considered.
The vertical dashed cuts (from left to right) correspond to $k_0\chi_0$ for gapped graphene (solid) at  
$\hbar\omega/2\Delta=0.18$, topological silicene in a quantum spin Hall insulator phase (dashed) at $\hbar\omega/2\Delta=0.5$ (see text), a 2DEG under a magnetic field $B_{\perp }=1\,$T (dotted), for $\hbar\omega/\hbar\omega_c=0.5$, and
bilayer graphene (dash-dotted) on hBN at $\hbar\omega/2\Delta=0.25$.}
\label{fig:Fig2}
\end{figure}

Figure\,\ref{fig:Fig2} shows the displacements $\delta^H_x$ (a) and $\delta^H_y$ (b), for incidence angle in an interval centered at the Brewster angle  $\theta_B\approx\arctan(\sqrt{\epsilon_2/\epsilon_1})=56.6^{\circ}$, 
defined by the reflection with $\sigma_{ii}=0$,
and as functions of the normalized static susceptibility $k_0\chi_0$. The range displayed in the horizontal axis includes the relevant span
of gapped Dirac-like systems, with
$k_0\chi_0\approx 10^{-4}$ at $\hbar\omega$ of few tens of meV ($\sim 1-10\,$THz).
As a first example, consider the model Hamiltonian 
$H_{\xi}({\bf k}) = \hbar v( \xi k_{x} \sigma_{x} + k_{y} \sigma_{y}) + \Delta \sigma_{z}$ ($\xi=\pm$ is a
valley index), for which Eq. (\ref{static2}) gives $\chi_0=g_sg_ve^2/12\pi|\Delta|$ \cite{Komissarov}
($g_s=2,\, g_v=2$ are the spin and valley degeneracies). At $\hbar\omega=20\,$meV  ($\sim 5\,$THz) 
the conductivity behaves linearly in frequency, and taking $\Delta=110\,$meV, we obtain $\chi_0=13.9$\AA\,(0.15\,aF in SI units) and
$k_0\chi_0=(\alpha/3\pi)(\hbar\omega/\Delta)=1.4\times 10^{-4}$ 
($\alpha=e^2/\hbar c$ is the fine structure constant).
Similar estimation is obtained for the model of a topological insulator described by a Dirac cone Hamiltonian
with the parabolic addition $-\hbar v\tilde{\alpha} k^2\sigma_z$ ($\tilde{\alpha}\Delta/\hbar v>0$) to the mass term $\Delta\sigma_z$ \cite{Komissarov}.
For a twisted bilayer graphene aligned on top of hBN, with a gap $2\Delta=4\,$meV \cite{LiuMolecules}, 
the value $\chi_0\approx 1528$\AA\,($17\,$aF) is obtained \cite{Komissarov}; at $\hbar\omega/2\Delta=0.25$,
$k_0\chi_0\approx 7.7\times 10^{-4}$. 
According to the generic low-energy Haldane model used in Ref.\,\cite{PSHE_Haldane} to study the 
optical conductivity and the PSHE in Haldane model materials, in the topological case we obtain $\chi_0=5.88$\AA\, (0.065aF) for typical parameters, such that $k_0\chi_0=0.6\times 10^{-4}$ at $\hbar\omega=20\,$meV. 
For a buckled Xene material like silicine, where the gap can be varied by electrical means and become
valley depending via spin-orbit coupling (inducing a gap of 1.55-7.9 meV), and using parameters for which a quantum spin Hall insulator
behavior is present \cite{PSHE_Haldane}, we estimate $\chi_0=522.5$\AA\,(5.8 aF) and $k_0\chi_0=5.3\times 10^{-4}$,
at $\hbar\omega=2\,$meV. In contrast, for a transition metal dichalcogenide monolayer like MoS$_2$, which presents
a larger band gap ($\sim 1.6\,$eV), the estimation is $\chi_0=1.84$\AA\,(0.02 aF) and $k_0\chi_0\approx 0.18-0.28\times 10^{-4}$ at $\hbar\omega=20-30\,$ meV.
In a 2D electron gas under an out-of-plane magnetic field $B_{\perp }=1\,$T, each
Landau level carries $\chi_0=19778$\AA\, (0.22\,fF) \cite{Komissarov}, such that for $\hbar\omega/\hbar\omega_c=0.5$, 
$k_0\chi_0=6\times 10^{-4}$ ($\omega_c=eB_{\perp}/m^*c$ is the cyclotron frequency; we use
$m^*=0.055m_0$). Some of these estimations are indicated by vertical dashed lines
in Fig.\,\ref{fig:Fig2}. For the sake of exhaustiveness, 
we extend the
range of plot up to $k_0\chi_0=4\times 10^{-3}$.
\begin{figure}[H]
\centering
\includegraphics[width=.45\textwidth]{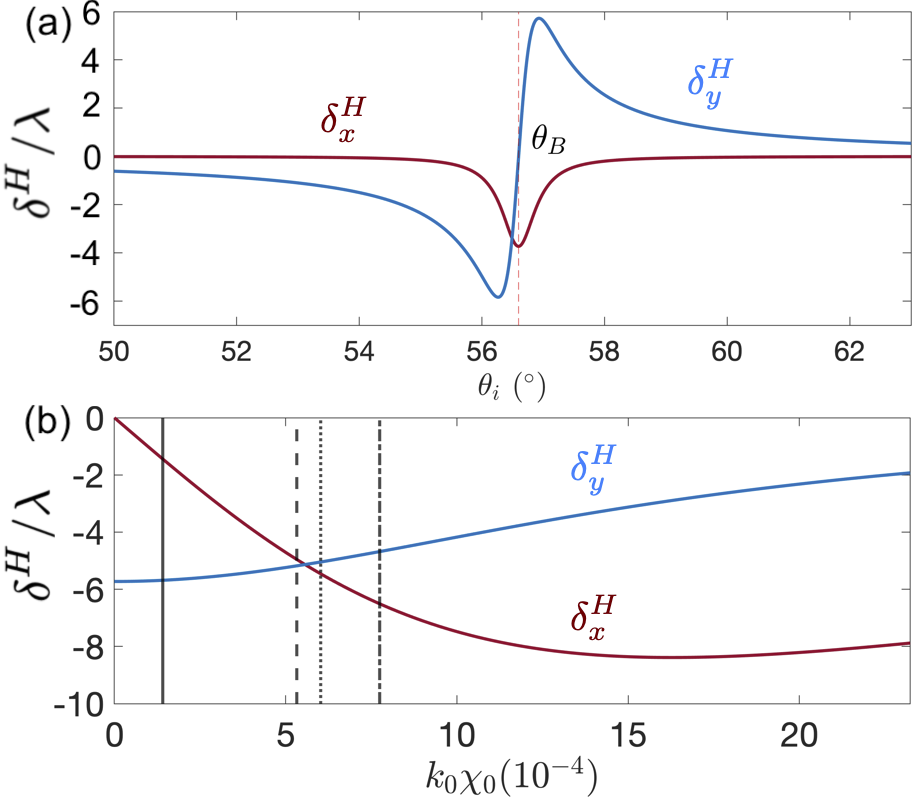}
\caption{(a) In-plane $\delta^H_x$ (red) and transverse $\delta^H_y$ (blue) shifts in the vicinity of the Brewster angle of incidence $\theta_B=\arctan(\sqrt{\epsilon_2/\epsilon_1})=56.6^\circ$,
for graphene with $\Delta=110\,$meV, at $\hbar\omega=40\,$meV and $k_0\chi_0=2.8\times 10^{-4}$
(solid line in Fig.\,\ref{fig:Fig2}). The Hall shift $\delta^H_y$
vanishes at an incidence angle very close to $\theta_B$. 
(b) Beam shifts as functions of $k_0\chi_0$ for
$\theta_i=0.9925\theta_B$. The vertical lines are the same as in Fig.\,\ref{fig:Fig2}.}
\label{fig:Fig3}
\end{figure}

It is noted in Fig.\,\ref{fig:Fig2} that, for $k_0\chi_0\sim 10^{-4}-10^{-3}$, the shift $\delta_x^H$ becomes significant for incidence around the angle $\theta_B$,
%$\theta_B\approx\arctan(\sqrt{\epsilon_2/\epsilon_1})=56.6^{\circ}$, 
presenting a dip of about 
$-10\lambda$ to $-12\lambda$ in size right at $\theta_i=\theta_B$ (Fig.\,\ref{fig:Fig2}(a)), while $\delta_y^H$ displays a change of sign at the same angle, with a dip and a peak of $\sim \pm 6\lambda$, symmetrically located in its vicinity (Fig.\,\ref{fig:Fig2}(b)).
This is illustrated in Fig.\,\ref{fig:Fig2}(a) where the beam shifts as functions of 
$\theta_i$ for gapped graphene are shown (the vertical solid line). 
The resonance-like structure between the shifts close to $\theta_B$
%(mimicking a Kramers-Kronig pair of a causal response function), 
arises
from the behavior (not shown) of $-\text{Im}(\partial \ln r_p/\partial\theta_i)$, which
displays a maximum, and Re($r_s/r_p$), which is negative for $\theta_i<\theta_B$
and positive for $\theta_i>\theta_B$.
Figure\,\ref{fig:Fig3}(b) shows the shifts for varying $k_0\chi_0$,
at fixed angle $\theta_i=0.9925\theta_B$.
Notably, they lie in the detectable range and should be possible to measure them with the
current tool based on weak-measurement technique \cite{Hosten,Qin2009}.

In the quasistatic regime behind Eq.~(\ref{static2}), the longitudinal conductivity satisfies \( |\sigma_{ii}|/c = k_0 \chi_0 \ll 1 \), and the Fresnel amplitude in Eq.\,(\ref{rpp}) can be approximated by $r_p=r_p^0[1-ia_1(k_0\chi_0)+a_2(k_0\chi_0)^2]$ with good accuracy for $\theta_i\neq\theta_B$, 
where $r_p^0(\theta_i)$ corresponds to the amplitude for the bare interface ($\sigma_{ii}=0$),
$a_1(\theta_i)=8\pi\epsilon_1k_{iz}k_{tz}^2/k_0[(\epsilon_{2} k_{iz})^{2} - (\epsilon_{1} k_{tz})^{2}]$
and $a_2(\theta_i)=4\pi k_{iz}k_{tz}a_1/k_0(\epsilon_2k_{iz}-\epsilon_1 k_{tz})$.
Using this approximation in Eq.\,(\ref{xHVpm}), we obtain
\begin{equation} \label{dxapprox}
\frac{\delta_x^H}{\lambda}\approx -\frac{1}{2\pi\Lambda^0_p}\frac{\partial a_1}{\partial\theta_i}\,k_0\chi_0, 
\end{equation}
to leading order in the parameter $k_0\chi_0$,
where $\Lambda^0_p$ is the quantity (\ref{denominator}) after taking $\sigma_{ii}=0$.
The terms omitted in Eq.\,(\ref{dxapprox}) involve increasing odd powers of $k_0\chi_0$ only.
In contrast, for the transverse shift the expansion of Eq. (\ref{xHVpm}) gives
\begin{equation} \label{dyapprox}
\frac{\delta_y^H}{\lambda}\approx \frac{1}{2\pi}\frac{\cot\theta_i}{\Lambda^0_p}
\left[1 + \frac{r_s^0}{r_p^0}+b_2(k_0\chi_0)^2\right], 
\end{equation}
where $r^0_s$ for $s$ polarization is defined similarly to $r^0_p$.
The expression for the coefficient $b_2(\theta_i)$ 
is long and we omit it for brevity. The next terms in the expansion involve increasing even powers
of $k_0\chi_0$ only.

\begin{figure}[H]
\centering
\includegraphics[width=.4 \textwidth]{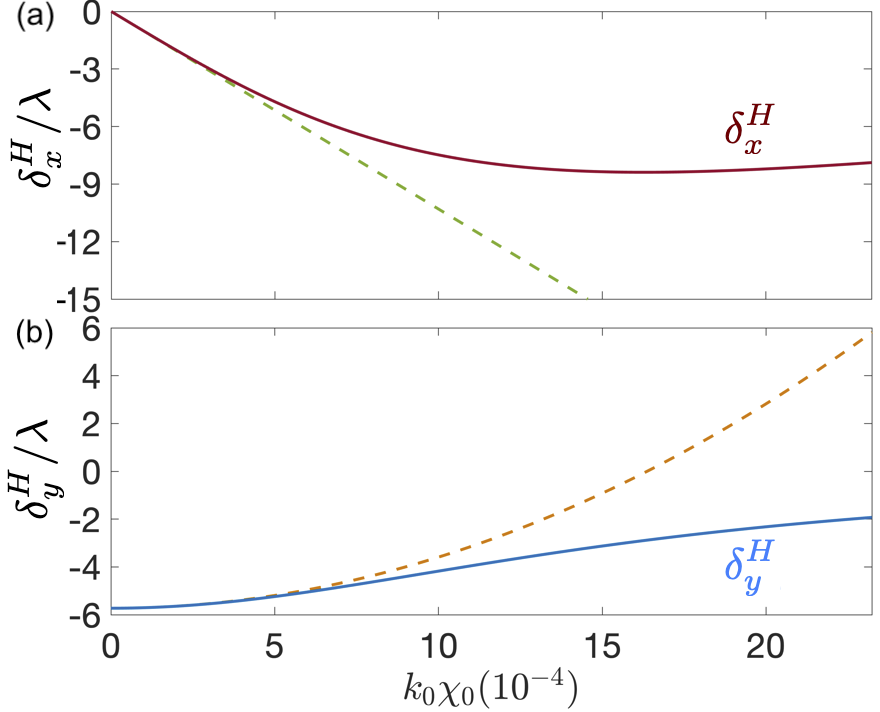}
\caption{Polynomial approximations (dashed lines) of the in-plane (a) and transverse (b) beam shifts for
varying normalized capacitance $k_0\chi_0$ at fixed angle $\theta_i=0.9925\theta_B$, 
compared to the exact (numerical) results (solid lines). In (a) the approximation is linear in $k_0\chi_0$ 
while in (b) is quadratic (Eqs. (\ref{dxapprox}) and (\ref{dyapprox})).}
\label{fig:Fig4}
\end{figure}

Figure \ref{fig:Fig4} displays these approximations compared to the 
exact numerical results obtained from Eqs. (\ref{xHVpm}) and (\ref{yHVpm}). A good
agreement is observed for low enough frequencies. These approximations suggest that after measurements of the shifts,
Eqs. (\ref{dxapprox}) or (\ref{dyapprox}) could be used to estimate the quantum geometric susceptibility 
$\chi_0$. This can be useful to compare with the estimation obtained from the expression in Eq.(\ref{static2}) after quantum metric measurement.

An upper bound for $\delta_x^H$ can be obtained from Eq.\,(\ref{dxapprox}) and the inequality implied by Eq. (\ref{static2}), $\chi_0\leqslant (e^2/\varepsilon_{gap})g$, where
$g=\sum_{n,m\neq n}\int[d^2k/(2\pi)^2]f_{n{\bf k}}(1-f_{m{\bf k}})(g^{xx}_{nm}+g^{yy}_{nm})$ 
is the quantum metric of the occupied band manifold and $\varepsilon_{gap}$ denotes the energy gap,
\begin{equation}
\frac{\delta_x^H}{\lambda}\leqslant -\frac{\alpha}{2\pi}\frac{1}{\Lambda_p^0}\frac{\partial a_1}{\partial \theta_i}
\left(\frac{\hbar\omega}{\varepsilon_{gap}}\right)g\,.
\end{equation}
In a similar way, an upper bound for $\delta^H_y$ can be found. 
These inequalities enable the estimation of a lower bound for the quantum metric of 2D band insulators from low-frequency beam shift measurements.
Note that the prefactor of $(\hbar\omega/\varepsilon_{gap})g$ depends only on the incidence angle $\theta_i$ and the dielectric constants $\epsilon_1$ and $\epsilon_2$.

The temperature dependence of the optical conductivity will have an effect on the
beam shifts. To explore finite-temperature effects we perform a calculation for a model
Hamiltonian of a single Dirac cone with a gap. We find from Eq.\,(\ref{static2}) that for
$k_BT\ll\Delta$, $\chi_0$ is modified to $\chi_0f(T)$, where $\chi_0=e^2/12\pi|\Delta|$ and
$f(T)=\sinh(\Delta/k_BT)/[\cosh(\Delta/k_BT)+\cosh(\varepsilon_F/k_BT)]$. 
For $k_BT\lesssim 20\,$meV,
$\Delta\approx 100\,$meV (gapped graphene), and $\varepsilon_F=20\,$mev,
the factor $f(T)\approx 1$, suggesting that the low frequency PSHE is a robust effect even close to the room temperature.
Notwithstanding, the issue deserves a more exhaustive investigation.

In summary, we have studied the photonic spin Hall effect (PSHE) in the quasistatic (longitudinal) response regime of two-dimensional semiconductors and insulators. This low-frequency, non-dissipative response is determined by a static susceptibility, which can be understood as a quantum capacitance of geometric origin due to its relation to the quantum metric tensor. Using typical values of the normalized (dimensionless) capacitance \(\omega\chi_0/c\) for graphene-family materials, which are on the order of \(10^{-4}\), we calculated the in-plane and transverse beam shifts within the range \(\omega\chi_0/c = 0\) to \(10^{-3}\). We found that these shifts can reach magnitudes between \(\lambda\) and \(10\lambda\) at sufficiently low frequencies. In addition, we derived polynomial approximations that may be used to estimate the quantum geometric capacitance from displacement measurements, and to establish upper bounds for the shifts or lower bounds for the quantum metric.

%solo cambio de redacción, original abajo
Our findings align with recent studies on static screening from interband transitions in doped graphene and phosphorene, which highlight the relevance of THz spectroscopy for probing 2D materials~\cite{modeling}. Potential applications of the PSHE in the THz regime have also been proposed for monolayer black phosphorus~\cite{PSHE_BPh} and in magneto-optical graphene-substrate systems~\cite{MO-PSHE}. The sensitivity of the PSHE to topological phase transitions in staggered 2D semiconductors such as silicene has been demonstrated within the THz range, even away from optical transition frequencies~\cite{MShah,Kort-Kamp}, and optical detection of these transitions through enhanced in-plane PSHE signals using weak-value amplification has recently been reported~\cite{weak-value}. In this context, we propose the THz PSHE as a sensitive probe of the quantum geometric capacitance in 2D band insulators. We hope that our results will motivate further theoretical and experimental developments in this direction.

The authors acknowledge financial support of DGAPA-UNAM-PAPIIT project IN112125, México.
Y.F.M. acknowledges financial support from SECIHTI M\'exico.

\bibliography{biblio.bib}

\end{document}